\documentclass[aps,preprint,superscriptaddress,showpacs,floatfix]{revtex4}%
\usepackage{amsfonts}
\usepackage{amsmath}
\usepackage{amssymb}
\usepackage{graphicx}
\usepackage{graphicx}%
\begin{document}
\title{Transformation of subradiant state to superradiant state in a thick resonant medium}
\author{R. N. Shakhmuratov}
\affiliation{Kazan Physical-Technical Institute, Russian Academy of Sciences, 10/7 Sibirsky
Trakt, Kazan 420029 Russia}
\affiliation{Kazan Federal University, 18 Kremlyovskaya Street, Kazan 420008 Russia}
\pacs{42.25.Bs, 42.50.Gy, 42.50.Nn}
\date{{ \today}}

\begin{abstract}
The propagation of a step pulse through a thick resonant absorber with
homogeneously broadened absorption line is considered. It is shown that a
specific subradiant state is naturally developed in the absorber due to the
formation of the spatial domains of the atomic coherence with opposite phases.
It is proposed to divide the absorber into slices in accord with these domains
and place the phase shifters in front of the first slice and between the other
slices. If the phase shifters are switched on simultaneously at a particular
moment of time, elapsed from the beginning of the step pulse, a strong sharp
pulse is generated at the output of the last slice of the absorber. The effect
is explained by the phasing of the atomic coherence along all slices of the
absorber, which transforms subradiant state of atom-field system to
superradiant state.

\end{abstract}
\maketitle

\section{Introduction}

Resonant interaction of a weak coherent pulse with extended dielectric
absorbers in a linear regime is well described by the exact integral
representation of the propagating planewave field if the spectrum of the input
pulse and the complex dielectric permittivity of the medium are known (see,
for example, Refs. \cite{Crisp,Oughstun}). This integral is quite simple and
can be easily calculated numerically. Meanwhile, sometimes it is not obvious
how to calculate the output filed if we modify the field or the induced
polarization inside the medium during the pulse propagation. In this case
knowledge of the Green's function of the dispersive dielectric medium (its
response to a $\delta$-function input pulse) helps to find the shape of the
output pulse (see, for example, Ref. \cite{Yablonovitch}).

Change of the field phase or change of the induced polarization in the medium
along the pulse propagation, are quite effective methods of creating
subradiant and superradiant states. Transferring the radiation field between
these states was proposed in sequel of papers (see Refs.
\cite{Kalachev1,Kalachev2,Kalachev3,Kalachev4,Kalachev5}) to create a quantum
memory for single photons. In these papers such a transformation was intended
to implement by coherent pulses exciting atoms on auxiliary transitions
(adjacent to a resonant one), by a particular variant of CRIB technique
(Controlled Reversible Inhomogeneous Broadening), or by controllable phase
modulators of the field (phase shifters), inserted into the atomic ensemble in
a regular way along the direction of the signal pulse propagation. All these
methods imply an artificial creation of periodic spatial domains from atomic
dipoles, induced by the input signal field, such that dipoles in neighboring
domains have opposite phases. This operation produces a locked atom-field
subradiant state. It is a storage stage in a quantum memory protocol. In a
reading stage all domains of polarization are brought in phase, which results
in the superradiance, i.e., fast release of the radiation field.

In this paper it is found that a step (or rectangular) pulse propagating in an
optically thick resonant medium creates such domains of polarization with
opposite phases naturally. Their lengths are not equal and evolve in time.
Thus, we may take for granted a naturally built up atom-field subradiant
state. It is proposed to cut the sample into unequal slices in a particular
way and place phase shifters between them. Then at a given moment of time
their fast switch on results in the superradiance seen as a short and strong pulse.

\section{Spatial oscillations of the atomic polarization along the light beam
in a thick resonant medium}

For simplicity we consider the propagation of a weak pulse in a thick resonant
medium with homogeneously broadened absorption line. A generalization to the
case of inhomogeneously broadened absorption line with Lorenzian shape is
trivial (see, for example, Ref. \cite{Crisp}).

In the slowly varying amplitude (SVA) approximation the unidirectional
propagation of a weak pulse, $E(z,t)$, as a plane wave along axis $\mathbf{z}$
is described by a couple of the atom-field equations appropriate for the
linear response (LR) approximation. These equations are
\begin{equation}
\dot{\sigma}_{eg}=-\gamma\sigma_{eg}+i\Omega(z,t), \label{Eq1}%
\end{equation}%
\begin{equation}
\widehat{L}\Omega(z,t)=i\alpha\gamma\sigma_{eg}(z,t)/2, \label{Eq2}%
\end{equation}
where $\rho_{eg}=\sigma_{eg}\exp(-i\omega t+ikz)$ is the nondiagonal element
of the atomic density matrix and $\sigma_{eg}$ its slowly varying part;
$\omega$ and $k$ are the frequency and wave number of the pulse propagating
along axis $\mathbf{z}$ (here $z$ is a coordinate along $\mathbf{z}$);
$\gamma$ is the decay rate of the atomic coherence; $\Omega(z,t)=d_{eg}%
E_{0}(z,t)/2\hbar$, $d_{eg}$ is a matrix element of the dipole transition
between ground, $g$, and excited, $e$, states of an atom; $E_{0}%
(z,t)=E(z,t)\exp(i\omega t-ikz)$ is a slowly varying field amplitude;
$\widehat{L}=\partial_{z}+c^{-1}d_{t}$, and $\alpha$ is the resonant
absorption coefficient. We limit our consideration to the case of exact resonance.

These equations are easily solved by means of the Fourier transform,
\begin{equation}
F(\nu)=\int_{-\infty}^{+\infty}f(t)e^{i\nu(t-t_{0})}dt. \label{Eq3}%
\end{equation}
which reduces them to a couple of algebraic equations%
\begin{equation}
\sigma_{eg}(z,\nu)=-\frac{\Omega(z,\nu)}{\nu+i\gamma}, \label{Eq4}%
\end{equation}%
\begin{equation}
\left[  \frac{\partial}{\partial z}-\frac{i\nu}{c}+A(\nu)\right]  =0,
\label{Eq5}%
\end{equation}
where%
\begin{equation}
A(\nu)=\frac{i\alpha\gamma/2}{\nu+i\gamma}. \label{Eq6}%
\end{equation}
The solution is
\begin{equation}
\Omega(z,\nu)=\Omega(0,\nu)\exp\left[  (i\nu z/c)-A(\nu)z\right]  ,
\label{Eq7}%
\end{equation}
where $\Omega(0,\nu)$ is the Fourier transform of the input field envelope at
the front face of the absorber with coordinate $z=0$. The inverse Fourier
transform of Eq. (\ref{Eq7}) gives the familiar expression for the development
of the radiation field in the resonant absorber with distance, that is%
\begin{equation}
\Omega(z,t)=\frac{1}{2\pi}\int_{-\infty}^{+\infty}\Omega(0,\nu)\exp\left[
-i\nu(t-z/c)-A(\nu)z\right]  . \label{Eq8}%
\end{equation}
Below for simplicity of notations we disregard small value $z/c$.

For the input step pulse, $\Omega_{\Theta}(0,t)=\Omega_{0}\Theta(t)$, which is
switched on at $t=0$ and has the amplitude $\Omega_{0}$ (here $\Theta(t)$ is
the Heaviside step function), the integral in the solution (\ref{Eq8}) was
calculated in Ref. \cite{Crisp}. The result is expressed in terms of infinite
sum of the Bessel functions of ascending order, multiplied by the
coefficients, depending on $\alpha$, $\gamma$ and $t$. To simplify calculation
of the integral in Eq. (\ref{Eq8}) it is usually reduced with the help of the
convolution theorem to (see, for example, Refs. \cite{Varoquaux,Macke,Segard}%
)
\begin{equation}
\Omega(z,t)=%
%TCIMACRO{\dint \nolimits_{-\infty}^{+\infty}}%
%BeginExpansion
{\displaystyle\int\nolimits_{-\infty}^{+\infty}}
%EndExpansion
\Omega(0,t-\tau)R(z,\tau)d\tau, \label{Eq9}%
\end{equation}
where $R(z,\tau)$ is the output radiation from the absorber of length $z$, if
the input radiation is a very short pulse whose shape is described by the
Dirac delta function, $\delta(t)$, i.e., $R(z,\tau)$ is the Green's function
of the absorber of thickness $z$. This function is
\cite{Burnham,Crisp,Varoquaux,Macke}
\begin{equation}
R(z,t)=\delta(t)-e^{-\gamma t}\Theta(t)\sqrt{\frac{b}{t}}J_{1}\left(
2\sqrt{bt}\right)  , \label{Eq10}%
\end{equation}
where $J_{1}\left(  x\right)  $ is the first-order Bessel function, and
$b=\alpha z\gamma/2$. The inverse value of $b$ is referred to as superradiant
time $T_{R}=1/b$ (see Ref. \cite{Kalachev5}). Then, the parameter $b$ may be
referred to as superradiant rate. It is also referred to as the effective
thickness parameter since for the absorber of thickness $z$ this parameter is
proportional to $z$.

For the input step pulse equation (\ref{Eq9}) is reduced to (see, for example,
Refs. \cite{Varoquaux,Sh12opt})
\begin{equation}
\Omega_{\Theta}(z,t)=\Omega_{0}\Theta(t)\left[  e^{-\gamma t}J_{0}\left(
2\sqrt{bt}\right)  +\gamma\int_{0}^{t}e^{-\gamma\tau}J_{0}\left(  2\sqrt
{b\tau}\right)  d\tau\right]  . \label{Eq11}%
\end{equation}
Another versions of this expression,%
\begin{equation}
\Omega_{\Theta}(z,t)=\Omega_{0}\Theta(t)\left[  1-\int_{0}^{t}e^{-\gamma\tau
}\sqrt{\frac{b}{\tau}}J_{1}\left(  2\sqrt{b\tau}\right)  d\tau\right]  ,
\label{Eq12}%
\end{equation}
can be found, for example, in Refs. \cite{Macke,Segard}). Meanwhile, the
equation for the output field can be expressed in a fast converging series
(see Appendix in Ref. \cite{Sh12opt}),%
\begin{equation}
\Omega_{\Theta}(z,t)=\Omega_{0}\Theta(t)\left\{  e^{-b/\gamma}+e^{-\gamma
t}\left[  f_{0}(b)J_{0}\left(  2\sqrt{bt}\right)  +\sum_{n=1}^{\infty}%
f_{n}(b,t)j_{n}(bt)\right]  \right\}  , \label{Eq13}%
\end{equation}
where $j_{n}(bt)=J_{n}\left(  2\sqrt{bt}\right)  /(bt)^{n/2}$, $J_{n}\left(
2\sqrt{bt}\right)  $ is the Bessel function of the $n$-th order,
$f_{0}(b)=f_{0}(b,t)=1-\exp(-b/\gamma)$, and%
\begin{equation}
f_{n}(b,t)=(\gamma t)^{n}\left[  1-e^{-b/\gamma}\sum_{k=0}^{n}\frac
{(b/\gamma)^{k}}{k!}\right]  , \label{Eq14}%
\end{equation}
It should be noted that the sum in Eq. (\ref{Eq14}) is a truncated Taylor
series for $\exp(b/\gamma)$. Thus, with increase of $n$ the expression in
square brackets in Eq. (\ref{Eq14}) tends to zero, i.e.,%
\begin{equation}
\lim_{n\rightarrow\infty}\left[  1-e^{-b/\gamma}\sum_{k=0}^{n}\frac
{(b/\gamma)^{k}}{k!}\right]  =0. \label{Eq15}%
\end{equation}

Depending on the values of the parameters $b$ and $\gamma$ it is enough to
take into account only one or two first terms in the sum, which is the third
term in Eq. (\ref{Eq13}), to obtain fine approximation for $\Omega_{\Theta
}(z,t)$. If $b\gg\gamma$, i.e., the superradiant rate is much faster than the
decay rate of the atomic coherence, then time evolution of the output field,
$\Omega_{\Theta}(z,t)$, is well approximated by the function%
\begin{equation}
\Omega_{\Theta}(z,t)\approx\Omega_{0}\Theta(t)e^{-\gamma t}J_{0}\left(
2\sqrt{bt}\right)  , \label{Eq16}%
\end{equation}
which is the main part of the second term in Eq. (\ref{Eq13}). Other terms
give minor contribution. This condition is satisfied if $\alpha l/2\gg1$,
where $l$ is the length of the absorbing sample.

To find the spatiotemporal evolution of the atomic coherence along the sample
we can use Eq. (\ref{Eq4}) and obtain according to the convolution theorem the
result%
\begin{equation}
\sigma_{eg}(z,t)=i\int_{-\infty}^{+\infty}e^{-\gamma(t-\tau)}\Theta
(t-\tau)\Omega(z,\tau)d\tau. \label{Eq17}%
\end{equation}
where $z$ changes between $0$, which is coordinate of the front face of the
sample, and $l$, which is coordinate of the sample end. For the step pulse Eq.
(\ref{Eq17}) is reduced to (see Ref. \cite{Sh12opt})
\begin{equation}
\sigma_{eg}(z,t)=i\Omega_{0}\Theta(t)\int_{0}^{t}e^{-\gamma\tau}J_{0}\left(
2\sqrt{b\tau}\right)  d\tau. \label{Eq18}%
\end{equation}
From Eqs. (\ref{Eq11}) and (\ref{Eq13}) it follows that the atomic coherence
can be expressed as%
\begin{equation}
\sigma_{eg}(z,t)=\frac{i\Omega_{0}}{\gamma}\Theta(t)\left\{  e^{-\gamma t}%
\sum_{n=1}^{\infty}f_{n}(b,t)j_{n}(bt)+e^{-b/\gamma}\left[  1-e^{-\gamma
t}J_{0}\left(  2\sqrt{bt}\right)  \right]  \right\}  . \label{Eq19}%
\end{equation}
If $b\gg\gamma$, then evolution of the atomic coherence is well approximated
by the main part of the first term in the sum in Eq. (\ref{Eq19}), which is
\begin{equation}
\sigma_{eg}(z,t)\approx i\Omega_{0}t\Theta(t)e^{-\gamma t}\frac{J_{1}\left(
2\sqrt{bt}\right)  }{\sqrt{bt}}. \label{Eq20}%
\end{equation}
The coherence $\sigma_{eg}(z,t)$ is pure imaginary since the radiation field
is in exact resonance. The sign of this coherence oscillates with time $t$ and
distance $z$ (since $b=\alpha z\gamma/2$) according to the Bessel function
$J_{1}\left(  2\sqrt{bt}\right)  $.

We choose time $t_{p}$ satisfying the condition $b_{l}t_{p}\gg1$, where
$b_{l}=\alpha l\gamma/2$ and $l$ is the coordinate of the output facet of the
sample. Spatial dependencies of the radiation field $\Omega_{\Theta}(z,t_{p})$
(dotted line) and imaginary part of the coherence $\sigma_{eg}(z,t_{p})$
(solid line) along the sample, if $\gamma t_{p}\rightarrow0$ and $b_{l}%
t_{p}=30$, are shown in Fig. 1. In this case $\Omega_{\Theta}(z,t_{p})$ and
$\sigma_{eg}(z,t_{p})$ are well described by Eqs. (\ref{Eq16}) and
(\ref{Eq20}), respectively. The plot of the field amplitude is normalized to
$\Omega_{0}$ and the atomic coherence to $\Omega_{0}t_{p}$, thus both are
defined as nondimensional values, which have the same maxima equal to $1$.

In one's mind the sample can be divided into several domains such that in the
neighboring domains the atomic coherences have opposite phases. This
phenomenon can be understood with the help of the concept of Feynman
\cite{Feynman} explaining how the light propagates in a linear regime through
a dielectric medium.

According to his concept the radiation field at the output of a finite size
sample can be considered as a result of the interference of the input field,
as if it would propagate without interaction, with the secondary field
radiated by the linear polarization induced in the sample. The secondary field
is actually a coherently scattered field whose phase is opposite to the phase
of the incident radiation field. Destructive interference of these fields
leads to attenuation of the radiation field at the output of the sample.

In Fig. 1 the sample is divided in three domains marked by vertical lines,
placed at coordinates $z$ (in units of $bt_{p}$) where $\sigma_{eg}(z,t_{p})$
is zero. Below we refer to the cordinates of the right borders of the domains
I, II, and III as $z_{1}$, $z_{2}$, and $z_{3}$, respectively.

In the domain I the imaginary part of the atomic coherence is positive.
Therefore, the phase of the coherently scattered field is opposite to the
phase of the incident field and the sum of these fields, $\Omega(z,t_{p})$, is
attenuated due to their destructive interference. As a result, the sum field
and atomic coherence decrease with distance.

At some distance the sum field, $\Omega(z,t_{p})$, becomes zero. However, this
process has some inertia due to the energy accumulation in the atomic
excitation. Therefore, further at a particular distance the scattered field
becomes even greater than the incident field, producing the phase change of
the sum field, $\Omega(z,t_{p})$. This is also confirmed by the wave equation
(\ref{Eq2}) rewritten in the retarded reference frame as%
\begin{equation}
\partial\Omega(z,t_{r})/\partial z=i\alpha\gamma\sigma_{eg}(z,t_{r})/2,
\label{Eq21}%
\end{equation}
where time is defined as $t_{r}=t-z/c$. From this equation it is seen that if
imaginary part of $\sigma_{eg}(z,t_{r})$ is positive, the spatial derivative
of the sum field is negative and this derivative becomes zero only if
$\sigma_{eg}(z,t_{r})=0$, which takes place at coordinate $z_{1}$. Thus,
before coordinate $z_{1}$ the atomic coherence forces to decrease the sum
field $\Omega(z,t_{p})$ and when the sum field becomes zero the atomic
coherence continues this tendency making the field amplitude negative.

After the point where the sum field becomes negative, the field in its turn
tends to reverse the phase of the coherence bringing its amplitude to zero at
$z_{1}$ (marked by the first grey circle on the left in Fig.1). This process
is oscillatory and repeated in the next domains.

According to Eqs. (\ref{Eq16}) and (\ref{Eq20}) (as well as to Eq.
(\ref{Eq21})) the absolute value of the sum field $\Omega(z,t_{p})$ reaches
its local maxima at coordinates where $\sigma_{eg}(z,t_{p})=0$. Since at these
points the amplitude of the coherently scattered field takes maximum values,
we propose to shift simultaneously the phase of the sum fields at the same
points by $\pi$, including the front face of the sample. Then we expect that
in all domains the incident and scattered fields will interfere constructively
producing a pulse of large amplitude. The position of the phase shifters (PS)
in the sample, cut into slices at coordinates $z_{1}$, $z_{2}$, and $z_{3}$,
is shown at the bottom of Fig. 1. Effectively such a phase shift of the sum
fields will force the atomic coherences to amplify the field along the whole
sample coherently, i.e., we will make effective phasing of the atomic
coherences in all domains shown in Fig. 1. The phases of the atomic coherences
in each domain will be $-\pi/2$ with respect to the fields incident to the domain.

\section{Domain I}

Assume that at time $t_{p}>0$ the value of the coherence of the particles,
located at the output facet of the sample, reaches its first zero value,
$\sigma_{eg}(l,t_{p})=0$. This condition is satisfied if $b_{l}t_{p}=3.67$. By
that time only the domain I of atomic coherence (see Fig. 1) is developed in
the sample. At the same time we instantly change the phase of the input field
by $\pi$. Then the incident field becomes in phase with the secondary
(coherently scattered) field. Their constructive interference should result in
a strong and short pulse.

To find the transients, induced by that phase shift, we consider the incident
radiation field $\Omega(0,t)$ as consisting of two pulses, i.e.,
\begin{equation}
\Omega_{1}(0,t)=\Omega_{\Theta}(0,t)-2\Omega_{\Theta}(0,t-t_{p}), \label{Eq22}%
\end{equation}
where $\Omega_{\Theta}(0,t)=\Omega_{0}\Theta(t)$ is the step pulse, defined in
the previous Section. Then, the output field is%
\begin{equation}
\Omega_{1}(z_{1},t)=\Omega_{\Theta}(z_{1},t)-2\Omega_{\Theta}(z_{1},t-t_{p}).
\label{Eq23}%
\end{equation}
Here the function $\Omega_{\Theta}(z_{1},\tau)$ is defined in Eqs.
(\ref{Eq11}) and (\ref{Eq13}) where $b=b_{1}=\alpha z_{1}\gamma/2$ and
$z_{1}=l$. If $b_{1}\gg\gamma$, the approximate equation (\ref{Eq16}) is
valid, and then
\begin{equation}
\Omega_{1}(z_{1},t)=\Omega_{0}\left[  \Theta(t)e^{-\gamma t}J_{0}\left(
2\sqrt{b_{1}t}\right)  -2\Theta(t-t_{p})e^{-\gamma(t-t_{p})}J_{0}\left(
2\sqrt{b_{1}(t-t_{p})}\right)  \right]  . \label{Eq24}%
\end{equation}
From this equation we see that just after $t_{p}$ ($t=t_{p}+0$) the amplitude
of the output field is%
\begin{equation}
\Omega_{1}(z_{1},t_{p})=\Omega_{0}\left[  e^{-\gamma t_{p}}J_{0}\left(
2\sqrt{b_{1}t_{p}}\right)  -2\right]  . \label{Eq25}%
\end{equation}
If $\gamma t_{p}\ll1$, then the maximum amplitude of the pulse is $\Omega
_{1}(z_{1},t_{p})=-2.4\Omega_{0}$, its phase is opposite to the phase of the
input field, and intensity is $5.76$ times larger than the intensity of the
incident radiation field. The shape of the pulse is shown in Fig. 2, where the
intensity of the output field is plotted without approximation (\ref{Eq16})
for different values of the decay rate of the atomic coherence $\gamma$.

\section{Domains I plus II}

In this section we consider the case if $b_{l}t_{p}=12.3$. Then, by time
$t_{p}$ two domains (I and II, see Fig. 1) are developed in the absorber. We
could cut the absorber of length $l$ into two slices of lengths $z_{1}$ and
$l-z_{1}$, where $z_{1}$ satisfies the relation $\alpha z_{1}\gamma
t_{p}/2=3.67$. Below we define the parameters $b_{1}$ and $b_{2}$ for these
slices, which are $b_{1}=3.67/t_{p}$ and $b_{2}=8.63/t_{p}$. Then, by
definition we have $b_{1}+b_{2}=b_{l}$. We can make a gap $\delta_{12}$
between these slices and place the phase shifters in fronts of each slice (see
the bottom of Fig. 1). When phase shifters are off the input and hence output
fields for the second slice acquire an additional phase factor $\exp
(ik\delta_{12})$ due to the gap between slices. Below we neglect this factor
since it does not affect the intensity of the output field from the composite absorber.

The output field from the first slice of the composite absorber, $\Omega
_{1}(z_{1},t)$, is described by Eq. (\ref{Eq23}), where the cooperative rate
$b$ is $b_{1}$. Due to the second phase shifter, switched on at $t_{p}$ (see
Fig. 1), the input field for the second slice is $\Omega_{1}(z_{1}%
,t)[1-2\Theta(t-t_{p})]$. The explicit expression for this field is%
\begin{equation}
\Omega_{\pi1}(z_{1},t)=\Omega_{\Theta}(z_{1},t)[1-2\Theta(t-t_{p}%
)]+2\Omega_{\Theta}(z_{1},t-t_{p}), \label{Eq26}%
\end{equation}
where index $\pi$ means the the phase of the field $\Omega_{1}(z_{1},t)$ is
changed by $\pi$. Here we disregard the distance $\delta_{12}$ between pieces
and put its value equal to zero since its contribution to the intensity of the
output field from the composite absorber is zero for any value of $\delta
_{12}$.

With the help of Eq. (\ref{Eq9}) one can calculate the radiation field,
$\Omega_{2}(z_{2},t)$, at the output of the second slice and obtain the
expression, which is reduced to
\begin{equation}
\Omega_{2}(z_{2},t)=\Omega_{\Theta}(z_{2},t)+2\Omega_{\Theta}(z_{2}%
,t-t_{p})-2\Theta(t-t_{p})\Omega_{12}(z_{1},z_{2},t,t_{p}).\label{Eq27}%
\end{equation}
Here the functions $\Omega_{\Theta}(z_{2},t)$ and $\Omega_{\Theta}%
(z_{2},t-t_{p})$ are defined in Eq. (\ref{Eq11}), where superradiant rate $b$
is substituted by $b_{l}=b_{1}+b_{2}$. These functions describe those
components of the output field from the second slice, which are produced by
the input fields $\Omega_{\Theta}(z_{1},t)$ and $\Omega_{\Theta}(z_{1}%
,t-t_{p})$, respectively. Their derivation is given in Appendix. The function
$\Omega_{12}(z_{1},z_{2},t,t_{p})$ (valid for $t\geqslant t_{p}$) is%
\begin{equation}
\Omega_{12}(z_{1},z_{2},t,t_{p})=\Omega_{\Theta}(z_{1},t)-b_{2}%
%TCIMACRO{\dint \limits_{0}^{t-t_{p}}}%
%BeginExpansion
{\displaystyle\int\limits_{0}^{t-t_{p}}}
%EndExpansion
\Omega_{\Theta}(z_{1},t-\tau)e^{-\gamma\tau}j_{1}(b_{2}\tau)d\tau.\label{Eq28}%
\end{equation}
It describes the transformation of the field $\Theta(t-t_{p})\Omega_{\Theta
}(z_{1},t)$ by the second slice of the absorber. The function $j_{1}(b_{2}%
\tau)$ is defined just after Eq. (\ref{Eq13}). Multiple integration in Eq.
(\ref{Eq28}) can be avoided if instead of Eq. (\ref{Eq11}) for $\Omega
_{\Theta}(z_{1},t-\tau)$ one uses Eq. (\ref{Eq13}).

Just after the phase shift of the fields (at time $t=t_{p}+0$) the amplitude
of the output field from the second slice is
\begin{equation}
\Omega_{2}(z_{2},t_{p})=\Omega_{\Theta}(z_{2},t_{p})+2\Omega_{\Theta}%
(z_{2},0)-2\Omega_{\Theta}(z_{1},t_{p}). \label{Eq29}%
\end{equation}
If $b_{1,2}\gg\gamma$, then according to the approximate equation (\ref{Eq16})
this amplitude is approximated as%
\begin{equation}
\Omega_{2}(z_{2},t_{p})=\Omega_{0}\left\{  2+e^{-\gamma t_{p}}\left[
J_{0}\left(  2\sqrt{(b_{1}+b_{2})t_{p}}\right)  -2J_{0}\left(  2\sqrt
{b_{1}t_{p}}\right)  \right]  \right\}  . \label{Eq30}%
\end{equation}
If $\gamma t_{p}\ll1$, then for the chosen values of the superradiant rates of
the slices, i.e., $b_{1}=3.67/t_{p}$ and $b_{2}=8.63/t_{p}$, the amplitude of
the output field is $3.1\Omega_{0}$ and its intensity is $9.65$ times larger
than the intensity of the incident radiation field. The shape of the pulse
appearing at the output of the composite absorber after the phase shift of the
fields is shown in Fig. 3 where the output field intensity is plotted
according to Eq. (\ref{Eq27}) for different values of the decay rate of the
atomic coherence $\gamma$ .

\section{Three domains}

If by time $t_{p}$ the superradiant rate of the absorber, $b_{l}$, satisfies
the relation $b_{l}t_{p}=25.87$, then three domains are developed inside the
absorber. We divide such absorber in three slices satisfying the relations
$b_{1}t_{p}=3.67$, $b_{2}t_{p}=8.63$, and $b_{3}t_{p}=13.57$, where
$b_{k}=\alpha l_{k}\gamma/2$, and $l_{k}$ is the thickness of $k$-th slice,
which is related to the coordinates $z_{k}$ of the domain borders (see Fig. 1)
as follows $l_{1}=z_{1}$, $l_{2}=z_{2}-z_{1}$, and $l_{3}=z_{3}-z_{2}$. We
make gaps between the slices and place phase shifters between them and before
the first slice.

In the previous Section it was shown that after $\pi$-phase shift of the
fields before the first and second slices the output field from the second
slice, $\Omega_{2}(z_{2},t)$, is described by Eq. (\ref{Eq27}). Due to its
$\pi$-phase shift, produced between the second and third slices by the phase
shifter, the input field for the third slice is%
\begin{equation}
\Omega_{\pi2}(z_{2},t)=\Omega_{\Theta}(z_{2},t)[1-2\Theta(t-t_{p}%
)]-2\Omega_{\Theta}(z_{2},t-t_{p})+2\Theta(t-t_{p})\Omega_{12}(z_{1}%
,z_{2},t,t_{p}). \label{Eq31}%
\end{equation}
With the help of Eq. (\ref{Eq9}) we calculate the amplitude of the radiation
field $\Omega_{3}(z_{3},t)$ at the output of the third slice, which is%
\begin{equation}
\Omega(z_{3},t)=\Omega_{\Theta}(z_{3},t)+2\Theta(t-t_{p}%
)[A(t)+B(t)+C(t)+D(t)], \label{Eq32}%
\end{equation}
where
\begin{equation}
A(t)=\Omega_{\Theta}(z_{1},t)-\Omega_{\Theta}(z_{2},t)-\Omega_{\Theta}%
(z_{3},t-t_{p}), \label{Eq33}%
\end{equation}%
\begin{equation}
B(t)=-b_{2}\int_{0}^{t-t_{d}}\Omega_{\Theta}(z_{1},t-\tau)e^{-\gamma\tau}%
j_{1}(b_{2}\tau)d\tau, \label{Eq34}%
\end{equation}%
\begin{equation}
C(t)=-b_{3}\int_{0}^{t-t_{d}}\left[  \Omega_{\Theta}(z_{1},t-\tau
)-\Omega_{\Theta}(z_{2},t-\tau)\right]  e^{-\gamma\tau}j_{1}(b_{3}\tau)d\tau,
\label{Eq35}%
\end{equation}%
\begin{equation}
D(t)=b_{2}b_{3}\int_{0}^{t-t_{d}}d\tau_{1}\int_{0}^{t-t_{d}-\tau_{1}}d\tau
_{2}\Omega_{\Theta}(z_{1},t-\tau_{1}-\tau_{2})e^{-\gamma(\tau_{1}+\tau_{2}%
)}j_{1}(b_{2}\tau_{2})j_{1}(b_{3}\tau_{1}). \label{Eq36}%
\end{equation}
Just after $t_{p}$ ($t=t_{p}+0$) the functions $B(t)$, $C(t)$, and $D(t)$ are
zero. Therefore, the amplitude of the output field at $t=t_{p}$ takes value
\begin{equation}
\Omega(z_{3},t_{p})=-2\Omega_{0}+2\Omega_{\Theta}(z_{1},t_{p})-2\Omega
_{\Theta}(z_{2},t_{p})+\Omega_{\Theta}(z_{3},t_{p}). \label{Eq37}%
\end{equation}
If $\gamma t_{p}\ll1$, then for the specified values of the parameters $b_{1}%
$, $b_{2}$, and $b_{3}$ the amplitude of the output field from the composite
absorber is $3.65$ times larger than the amplitude of the input field,
$\Omega_{0}$, and its intensity is $13.36$ times larger than the intensity of
the incident field. Thus, by the phase shift of the incident fields at the
inputs of the layers of the composite absorber we transform subradiant state
of the atom-field system to superradiant state, which is realized in emission
of a short and strong pulse. The shape of the pulse is shown in Fig. 4 for
different values of the decay rate of the atomic coherence.

\section{Discussion}

Recently the idea of the transformation of subradiant state to superradiant
state was experimentally verified with gamma-quanta propagating in the
sandwich absorbers \cite{Sh13}. Mechanical displacement of odd slices of the
composite absorber (sandwich) by a half-wavelength of the radiation field
allowed to phase the nuclear coherence along all slices. Since the wavelength
of gamma-quanta (86 pm for 14.4 keV photons) is extremely small such a
displacement was easily implemented by polyvinylidene-fluoride (PVDF)
piezopolymer thin film. In optical domain this method is inapplicable since
the radiation wavelength is three orders of magnitude larger. Therefore, in
this paper different method of effective phasing of the atomic coherence along
the composite absorber is proposed. In spite of this difference, comparison of
Eq. (\ref{Eq32}) with Eq. (45) in Ref. \cite{Sh13} shows some similarity of
the results. However, there is a qualitative difference in propagation of the
step pulse and exponentially decaying pulse through a thick resonant medium if
the decay rate of the latter is comparable with the decay rate of the atomic coherence.

It is also interesting to notice that the pulses, produced by the effective
phasing of the atomic coherence, look similar to the pulses, generated by
stacking of coherent transients \cite{Macke,Segard}. Physically they are
different since the stacking is produced by many pulses or phase switchings at
different moments of time. However the value of the maximum amplitude of the
pulse, generated, for example, by two phase switchings,%
\begin{equation}
\Omega_{\max}(t_{2})=\Omega_{0}\left\{  2-2e^{-\gamma t_{1}}J_{0}\left(
2\sqrt{bt_{1}}\right)  +e^{-\gamma t_{2}}J_{0}\left(  2\sqrt{bt_{2}}\right)
\right\}  , \label{Eq38}%
\end{equation}
almost coincides with the maximum amplitude, Eq. (\ref{Eq30}), generated from
the composite absorber consisting of two slices if $\gamma t_{1}\ll1$ and
$\gamma t_{2}\ll1$. The first term in Eq. (\ref{Eq38}) corresponds to the
transients induced by the second phase switching, the second term describes
transients induced by the first phase switching, and the last term describes
the transients induced by the leading edge of the step pulse. Here $b$ is the
superradiant rate of a single absorber, $t_{2}$ is a time interval between the
beginning of the step pulse and the second phase switching, $t_{1}$ is a time
interval between the first phase switching and the pulse generation. These
intervals a chosen such that the functions $J_{0}\left(  2\sqrt{bt_{1,2}%
}\right)  $ have local extrema. If many phase switchings of the field are
applied at particular moments of time, then, as estimated in Ref.
\cite{Macke,Segard}, the maximum intensity of the pulse is 156 times larger
than the intensity of the incident field. This is also applicable to the
multilayered absorber with a particular combination of thicknesses of the layers.

\section{Conclusion}

In this paper it is shown that during the step pulse propagation through a
thick resonant absorber the atomic coherence is formed into spatial domains
with opposite phases. As a result subradiant state is developed in the
absorber. It is proposed to divide the absorber into slices in accord with
these domains and place phase shifters between them and in the front of the
absorber. Fast phase switching of the incident fields at the input of each
slice transforms subradiant state to superradiant state seen as a strong and
short pulse.

\section{Acknowledgements}

This work was supported by National Science Foundation (Grant No. 0855668),
Russian Foundation for Basic Research (Grant No. 12-02-00263-a), and Program
of Presidium of RAS "Quantum mesoscopic and disordered systems".

\section{Appendix}

To calculate the output field from the second slice of the absorber if the
input field is $\Omega_{\Theta}(z_{1},t)$ or $\Omega_{\Theta}(z_{1},t-t_{p})$,
we use Eq. (\ref{Eq9}), which is reduced, for example, for $\Omega_{\Theta
}(z_{1},t)$ to the expression
\begin{equation}
\Omega_{\Theta}(z_{2},t)=\Omega_{\Theta}(z_{1},t)-\int_{0}^{t}\Omega_{\Theta
}(z_{1},t-\tau)e^{-\gamma\tau}\sqrt{\frac{b_{2}}{\tau}}J_{1}\left(
2\sqrt{b_{2}\tau)}\right)  d\tau. \label{EqA1}%
\end{equation}
Then, we apply the Laplace transform%
\begin{equation}
F(p)=\int_{0}^{+\infty}e^{-pt}f(t)dt, \label{EqA2}%
\end{equation}
to the function $\Omega_{\Theta}(z_{2},t)$. It can be done in the following
way. First, we calculate the Laplace transform of the function $\Omega
_{\Theta}(z_{1},t)$ in the form, represented in Eq. (\ref{Eq12}). The Laplace
transform of the function $\sqrt{b_{1}/t}J_{1}\left(  2\sqrt{b_{1}t}\right)  $
in the integral term in Eq. (\ref{Eq12}) can be found with the help of the
differentiation theorem. This Laplace transform is
\begin{equation}
1-e^{-b_{1}/p}. \label{EqA3}%
\end{equation}
Then, the Laplace transform of the whole integral term in Eq. (\ref{Eq12}) can
be calculated with the help of the linear transformation and integration
theorems. The result of this calculation is
\begin{equation}
\frac{1}{p}\left[  1-e^{-b_{1}/(p+\gamma)}\right]  . \label{EqA4}%
\end{equation}
Finally, the Laplace transform of $\Omega_{\Theta}(z_{1},t)$ is
\begin{equation}
\Omega_{\Theta}(z_{1},p)=\frac{1}{p}e^{-b_{1}/(p+\gamma)}. \label{EqA5}%
\end{equation}
Second, since the integral in Eq. (\ref{EqA1}) is the convolution of two
functions, the Laplace transform of the right hand side of Eq. (\ref{EqA1}) is%
\begin{equation}
\Omega_{\Theta}(z_{1},p)-\Omega_{\Theta}(z_{1},p)\left(  1-e^{-b_{2}%
/(p+\gamma)}\right)  . \label{EqA6}%
\end{equation}
Combining Eqs. (\ref{EqA5}) and (\ref{EqA6}), we obtain the Laplace transform
of $\Omega_{\Theta}(z_{2},t)$, which is%
\begin{equation}
\Omega_{\Theta}(z_{2},p)=\frac{1}{p}e^{-(b_{1}+b_{2})/(p+\gamma)}.
\label{EqA7}%
\end{equation}
Comparison of Eq. (\ref{EqA7}) with Eq. (\ref{EqA5}) gives the result that
transmission of the field $\Omega_{\Theta}(z_{1},t)$ through the second slice
of length $l-z_{1}$ just changes the argument $z_{1}$ of the function,
describing the field, to $z_{2}=l$. This result is obvious since without phase
shifters the light propagation through two slices, having effective thickness
parameters $b_{1}$ and $b_{2}$ and placed in a row, is the same as the light
propagation through the absorber with effective thickness parameter
$b_{l}=b_{1}+b_{2}$.

\newpage

\newpage

\begin{figure}[ptb]
\resizebox{0.6\textwidth}{!}{\includegraphics{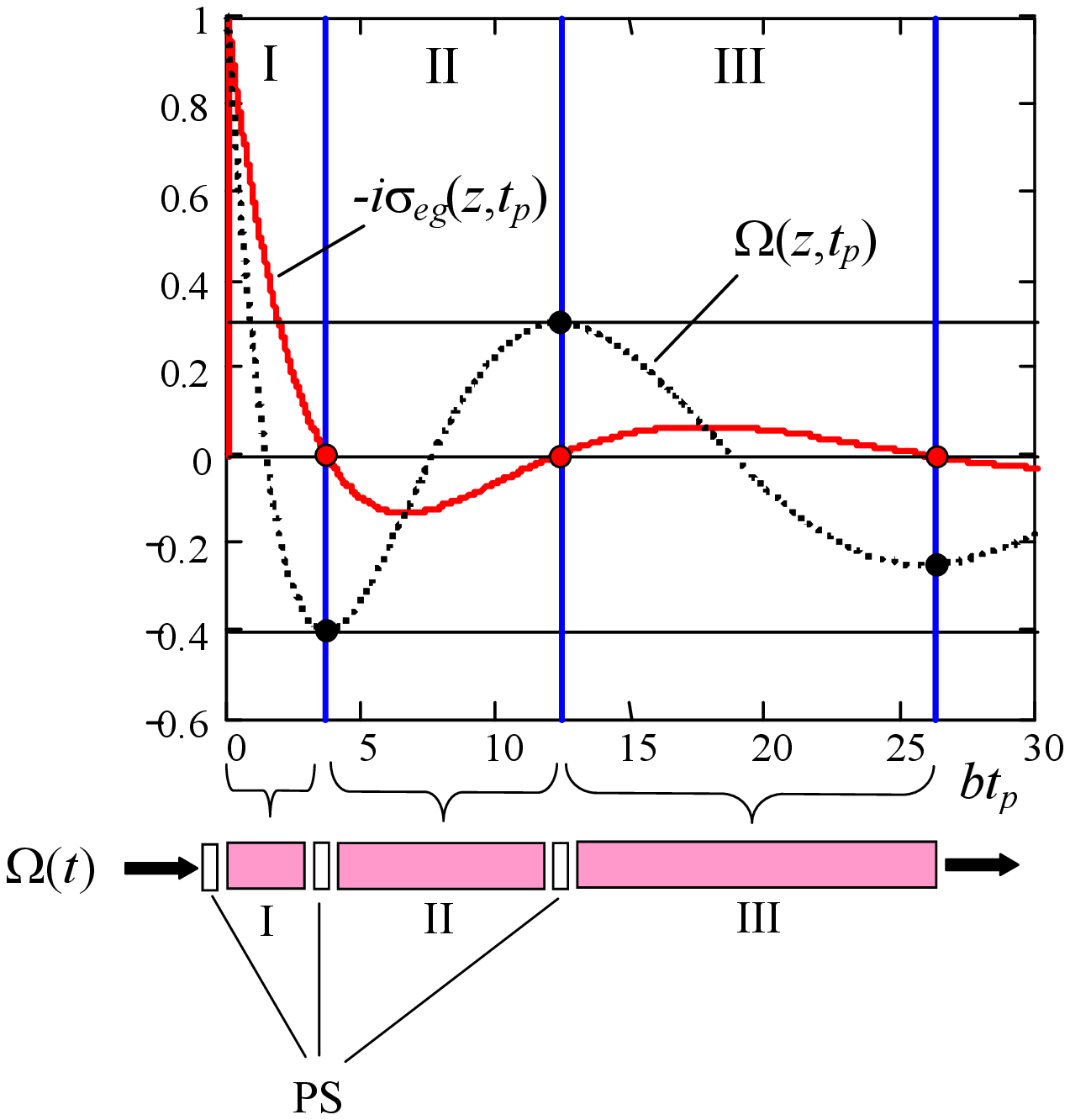}}\caption{(color on
line) Spatial dependencies of the imaginary part of the atomic coherence,
$-i\sigma_{eg}(z,t_{p})$, (solid line in red) and the radiation field,
$\Omega(z,t_{p})$, (dotted black line) along the absorber at $t=t_{p}$. Both
are normalized to 1 (see the text for details). Vertical bold blue lines
divide the plot into domains (I, II, and III), where the imaginary part of the
atomic coherence has one sign (plus or minus). Thin horizontal lines show the
values of the two first extrema of the field amplitude. The excitation scheme
of the absorber, cut into slices, and phase shifters (PS), placed between
them, are shown in the bottom.}%
\label{fig:1}%
\end{figure}

\newpage

\begin{figure}[ptb]
\resizebox{0.8\textwidth}{!}{\includegraphics{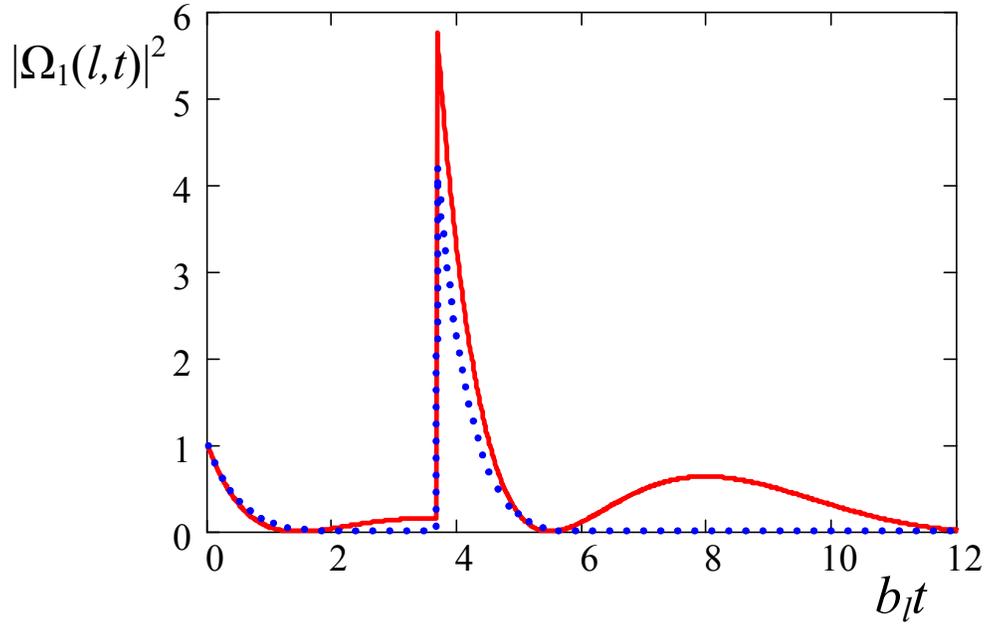}}\caption{(color on
line) The field intensity and the shape of the pulse, generated at the output
of the sample by the phase shift of the radiation field, applied at the input
at time $t_{p}$ satisfying the condition $b_{l}t_{p}=3.67$. $b_{l}$ is the
superradiant parameter of the sample. The field intensity is normalized to
$|\Omega_{0}|^{2}$. The decay rates of the atomic coherence are $\gamma
=0.003b_{l}$ (solid line in red) and $\gamma=0.3b_{l}$ (dotted line in blue).}%
\label{fig:2}%
\end{figure}

\newpage

\begin{figure}[ptb]
\resizebox{0.8\textwidth}{!}{\includegraphics{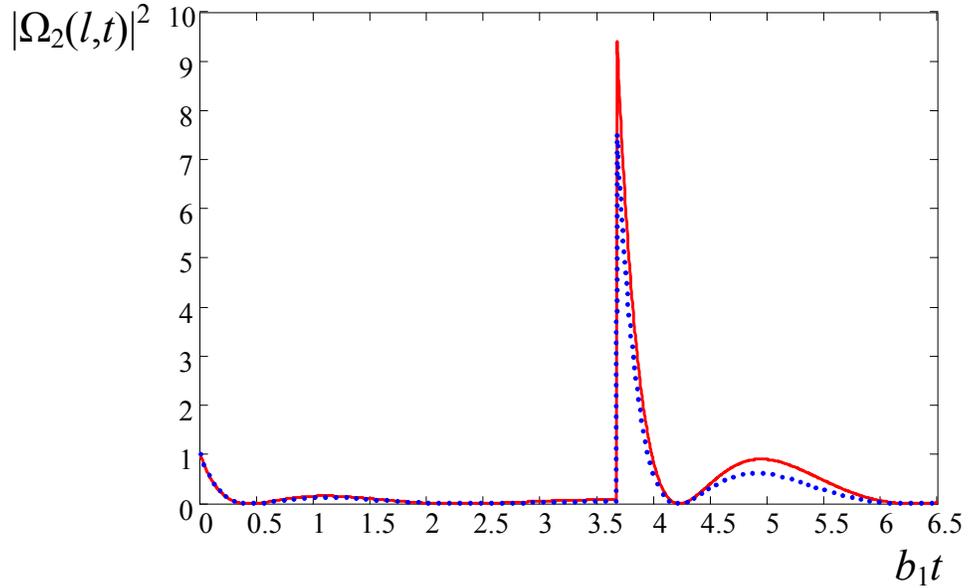}}\caption{(color on
line) The field intensity at the output of the composite absorber consisting
of two slices. The $\pi$-phase shift of the fields at the input of the first
and second slices of the composite absorber are applied at time $t_{p}$
satisfying the condition $b_{1}t_{p}=3.67$, where $b_{1}$ is the superradiant
rate of the first slice. Time scale is normalized to this rate. The field
intensity is normalized to $|\Omega_{0}|^{2}$. The decay rates of the atomic
coherence are $\gamma=0.01b_{1}$ (solid line in red) and $\gamma=0.1b_{1}$
(dotted line in blue).}%
\label{fig:3}%
\end{figure}

\newpage

\begin{figure}[ptb]
\resizebox{0.8\textwidth}{!}{\includegraphics{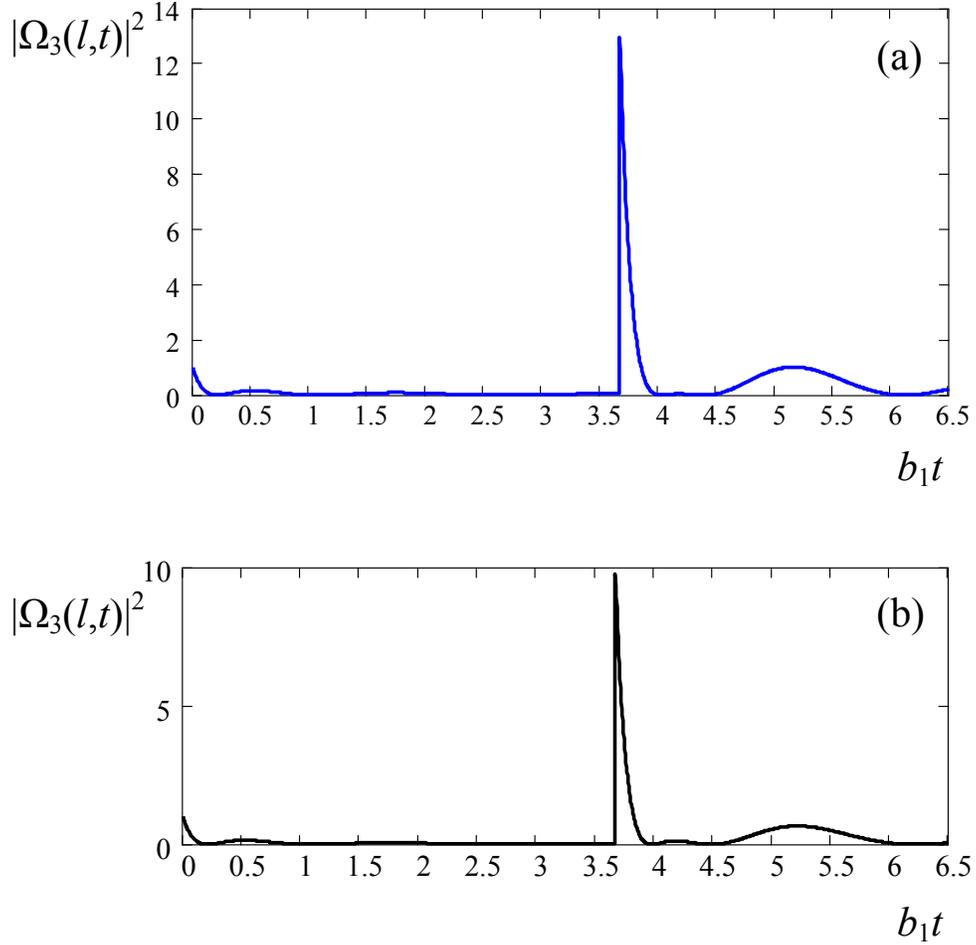}}\caption{(color on
line) The field intensity at the output of the composite absorber consisting
of three slices. The $\pi$-phase shift of the fields at the input of all
slices of the composite absorber are applied at time $t_{p}$ satisfying the
condition $b_{1}t_{p}=3.67$, where $b_{1}$ is the superradiant rate of the
first slice. Time scale is normalized to this rate. The field intensity is
normalized to $|\Omega_{0}|^{2}$. The decay rates of the atomic coherence are
$\gamma=0.01b_{1}$ (a) and $\gamma=0.1b_{1}$ (b).}%
\label{fig:4}%
\end{figure}

\end{document}